\documentclass[preprint]{jpsj2}
\title{
Mean-field model for magnetic
orders in NpTGa$_5$ with T=Co, Ni or Rh}

\author{Annam\'aria \textsc{Kiss}\thanks{E-mail address: amk@cmpt.phys.tohoku.ac.jp} and Yoshio
\textsc{Kuramoto}\thanks{E-mail address: kuramoto@cmpt.phys.tohoku.ac.jp}}

\inst{Department of Physics, Tohoku University, Sendai, 980-8578}

\abst{
Characteristics of  magnetic transitions in
NpTGa$_5$ with T=Co, Ni, Rh are explained in a unified way
with use of a crystalline electric field (CEF) model of localized
5$f^4$ electrons.
The model takes a CEF
doublet and a singlet as local states, and
includes dipolar and quadrupolar intersite interactions in the mean-field theory.
Diverse ordering phenomena are
derived depending on the
magnitude of interaction parameters, which qualitatively reproduce the
experimentally observed magnetic behaviors in NpTGa$_5$.
The quadrupole degrees of freedom are essential to the diverse magnetic orders.
It is argued that NpRhGa$_5$ is close to a multicritical point where
quadrupoles and dipoles with different directions are competing to order.
}

\kword{
actinide, Np, quadrupolar interaction, crystalline electric
field, magnetic order
}

\begin{document}
\maketitle

\section{Introduction}

Tetragonal $115$ compounds with $4f$ and $5f$ rare earth ions are
intensively studied recently because of their intriguing behavior.
For example,  superconductivity at $T_c = 18.5$ K has been
reported in PuCoGa$_5$ \cite{pu}. The $T_c$ is the highest among
the heavy-fermion superconductors. The second highest $T_c$ has
been observed in CeCoIn$_5$ at 2.3K\cite{petrovich} with the same
crystal structure. In heavy fermion systems the superconductivity
and magnetism can coexist, and their possible interplay is a
longstanding problem. Recently, NpTGa$_5$ systems with various
transition metal ions T attract much attention because of their
diverse magnetic behavior. Among these systems, NpCoGa$_5$ shows
an antiferromagnetic (AFM) phase transition at $47$K, which can
clearly be seen as a peak in the magnetic
susceptibility\cite{aoki,metoki, colineau}. It was found by
neutron diffraction that the ordered moments are parallel to the
tetragonal $c$-axis, and they have the in-plane ferromagnetic
structure, while the interplane stacking is antiferromagnetic.
Applying magnetic field to this system the ordered phase is
suppressed and metamagnetic transition occurs at low temperatures.
There are two magnetic transitions in both NpNiGa$_5$ and
NpRhGa$_5$, but the magnetic structures and the nature of the
ordered states are very different in these two systems. The
magnetic structure of NpRhGa$_5$ is the same as that in NpCoGa$_5$
below the first transition, but with further decrease of
temperature the moments switch over to lying within the $ab$-plane
by a first order transition\cite{jonen,aoki2,colineau2}. The
discontinuous transition is shown as a jump in the magnetic
susceptibility and a sharp increase of the magnetic scattering
intensities\cite{jonen}. In NpNiGa$_5$, a canted AFM structure is
found at low temperatures evolving from a ferromagnetically
ordered phase of moments lying parallel to the
$c$-axis\cite{honda}. In this compound both transitions are
continuous as indicated by finite jumps in the specific heat. At
the lower phase transition a clear anomaly can be seen in the
temperature dependence of the total magnetic moment\cite{honda2}.
Perpendicular magnetic order with moments lying within the
$ab$-plane is realized in NpFeGa$_5$ below the N\'eel temperature
118K. A further weak anomaly was found recently in the
thermodynamic quantities such as specific heat and magnetic
susceptibility at a lower temperature $\sim 78$K, which indicates
a change of the magnetic structure\cite{aoki2}. The presence of
magnetic moments at Fe sites makes the behavior more complex in
this compound.

In this paper, we explain the mechanism of diverse magnetic
orderings of NpTGa$_5$ systems on the basis of localized picture
of 5$f$ electrons. The observed effective magnetic moment in the
high-temperature part of the susceptibility in NpCoGa$_5$ is
consistent with the Np$^{3+}$ ($5f^4$) configuration, but it
highly deviates from the value for Np$^{4+}$ ($5f^3$) \cite{aoki}.
Furthermore, the magnetic properties of this system above the AFM
transition temperature is consistent with a low-lying
doublet--singlet crystal field level scheme, where the doublet is
the ground state. The next CEF level above this quasi-triplet is
lying at about 1200K \cite{aoki}. Therefore, the low-temperature
physics should mainly be determined by the pseudo-triplet states.
Due to the layered structure of 115-tetragonal systems, the most
important interactions are within the $ab$-planes. We introduce a
two-dimensional model by taking the doublet--singlet CEF model
within the 5$f^4$ configuration to explain the complex properties
of NpTGa$_5$ with T=Co, Ni and Rh. The quadrupole degrees of
freedom are found to be essential in understanding the diverse
magnetic behavior of NpTGa$_5$ systems. 

This paper is organized as follows. In section~2 we introduce the
model Hamiltonian within the pseudo-triplet subspace by clarifying
the relevant multipolar interactions. Three different limits of
the model are studied in details in section~3, and properties of
compounds with T=Co, Rh and Ni are explained in connection with
these limits. The summary and discussion is given in the last
section.

\section{Doublet--singlet CEF model}


The Hund's rule ground state of $5f^4$ configuration of Np$^{3+}$
ions is $L=6$ and $S=2$, which gives $J=4$ as the total angular
momentum. In tetragonal symmetry the nine-fold degenerate $J=4$
multiplet splits into five singlet and two doublet states. We work in
the following doublet-singlet local Hilbert space
\begin{eqnarray}
|d_{\pm}\rangle &=&a|\pm 3\rangle+\sqrt{1-a^2}|\mp 1\rangle\nonumber\\
|s \rangle &=&c\left(|+ 4\rangle+|- 4\rangle\right)+\sqrt{1-2c^2}|
0\rangle\,,\label{eq:qt}
\end{eqnarray}
which seems to be consistent with the magnetic properties of NpCoGa$_5$ in the paramagnetic phase\cite{aoki}.
We introduce $\Delta$ as
the energy separation between the doublet ground state and the singlet state. The decomposition
\begin{equation}
(E+A_1)\otimes (E+A_1) =
 2A_{1g}\oplus A_{2u}\oplus B_{1g}\oplus B_{2g}\oplus E_{g}\oplus
E_{u}\nonumber
\end{equation}
shows that the pseudo-triplet space carries the $J_z$ ($A_{2u}$)
and [$J_x$, $J_y$] ($E_{u}$) dipoles and $O_{2}^2$ ($B_{1g}$),
$O_{xy}$ ($B_{2g}$), [$O_{zx}$, $O_{yz}$] ($E_{g}$) quadrupoles as possible local order parameters.


The most important interactions are within a two-dimensional layer including Np ions.
For each $ab$-plane,
the dipole operators at site $i$ are written as $J_{k,i}$ with $k=x,y,z$, and
quadrupole operators as $O_{\mu,i}$ with $\mu=yz,zx,xy$.
We consider the nearest-neighbor interactions given by
\begin{eqnarray}
{\mathcal H}_{ab} =
-\frac 14\sum_{\langle ij\rangle} \left(  \sum_k
\Lambda_k J_{k,i}  J_{k,j}
+\sum_\mu
\Lambda_\mu O_{\mu,i} O_{\mu,j}  \right),
\label{eq:ham101}
\end{eqnarray}
where $\langle ij\rangle$ is a nearest-neighbor pair within a
tetragonal $ab$--plane. We take the mean-field theory where the
factor $1/4$ accounts for the number of nearest-neighbors.

Taking the CEF parameters in basis (\ref{eq:qt}) as $a=0.87$ and
$c=0.48$ seems appropriate to describe the high-temperature
magnetic properties of NpCoGa$_5$\cite{aoki}. 
In the absence of further information, we use these parameters also in the cases of T=Ni and Rh. 
With these CEF parameters
the eigenvalues of operators $J_z$, $J_x$, $O_{zx}$ and $O_{xy}$
are derived as $\pm 2.03$, $\pm 1.98$, $\pm 2.35$ and $\pm 2.19$,
respectively. Since all the eigenvalues are around $\pm 2$, we
normalize the operators $J_{k}$ and $O_{\mu}$ to the same value
for simplicity of calculation, and for transparency of the model.
Namely, we put their eigenvalues to be $\pm 1, 0$ within the
pseudo-triplet subspace so that the interaction parameters should
roughly be multiplied by 1/4 for the estimate of their magnitude.

The interaction Hamiltonian ${\mathcal H}_{ab} $ leads to very
complex ordering phenomena even within the pseudo-triplet
subspace.  We study the following limiting cases of the model. As
the simplest limit, we include only $\Lambda_{z}$ and
$\Lambda_{xy}$ as the intersite interactions (Case~I). Then the
singlet excited state and the doublet are decoupled. Next, we
introduce nonzero dipole interaction $\Lambda_x =\Lambda_y$
keeping the quadrupolar interaction $\Lambda_{xy}$ (Case~II).
Finally, we keep both dipole interactions $\Lambda_x =\Lambda_y$
and $\Lambda_z$, but we assume that the quadrupolar interactions
$\Lambda_{zx} =\Lambda_{yz}$ dominate over $\Lambda_{xy}$
(Case~III). We argue that these limits are relevant to describe
qualitatively the main properties of compounds with Co, Rh and Ni,
respectively. Table~\ref{tab:1} summarizes the limiting cases.
We consider the two-dimensional model given by (\ref{eq:ham101}) in Case~II and Case~III, while only in Case~I we additionally introduce Ising-type interlayer coupling in order to calculate also properties in external magnetic field.

\begin{table}
\begin{center}
\caption{Limiting cases of the model.
Dominant components of dipole and quadrupolar interactions are shown in each case.}
\label{tab:1}

\medskip
\begin{tabular}{c|c|c|c}
\hline  & dipole & quadrupole & relevance\\
\hline \hline Case I & $\Lambda_z$  &  $\Lambda_{xy}$ & NpCoGa$_5$ \\
\hline Case II & $\Lambda_z$, $\Lambda_x=\Lambda_y$ & $\Lambda_{xy}$ & NpRhGa$_5$ \\
\hline Case III & $\Lambda_z$, $\Lambda_x=\Lambda_y$ & $\Lambda_{zx}=\Lambda_{yz}$ &  NpNiGa$_5$\\
\hline
\end{tabular}
\end{center}
\end{table}

\section{Limiting cases of the model}

\subsection{Case~I}

As the simplest limit of the model,  we first consider only $\Lambda_z$ and $\Lambda_{xy}$ setting $\Lambda_x=\Lambda_y=\Lambda_{zx}=\Lambda_{yz}=0$. Then the singlet excited state is decoupled.
The relevant operators in the doublet are represented by the following matrices:
\begin{eqnarray}
J_z = \left(
\begin{array}{cc}
         1  & 0 \\
           0 & -1 \end{array}
       \right), \hspace*{0.5cm}
O_{xy}= \left( \begin{array}{cc}
         0  & -i \\
           i & 0 \end{array}
       \right).
\label{eq:doublet}
\end{eqnarray}
Thus the doublet can be diagonalized in two different ways leading to
Ising-like magnetism ($J_z$)
or quadrupolar order ($O_{xy}$).
The ground state is magnetically ordered with
$\langle J_z\rangle\ne 0$ for $\Lambda_z/\Lambda_{xy}>1$,
while for $\Lambda_z/\Lambda_{xy}<1$
we get quadrupole-ordered ground state with $\langle O_{xy}\rangle\ne 0$.
If we neglect the interlayer interaction,
the dipole transition temperature $T_z$ is simply given by
$T_{z}=\Lambda_z$, while
the quadrupole transition temperature $T_Q$ is given by
$T_{Q}=\Lambda_{xy}$.

We argue that the situation in NpCoGa$_5$, which has a single
phase transition at $T_{h\rm N}=47$K, can be described by the
doublet limit with $\Lambda_z>\Lambda_{xy}$. The energy separation
$\Delta \sim 87$K, as estimated as from the high-temperature
susceptibility, should be important to the susceptibility
anisotropy for example. However, the singlet does not influence
the main properties of the model such as the pattern of ordered
phases or the nature of the phase transitions.

The AFM ordering in NpCoGa$_5$ is characterized by the ordering
vector ${\bf K}=(0,0,1/2)$ in unit of $(2\pi/a,2\pi/a,2\pi/c)$. In
the following, ordering vectors ${\bf k}$ or ${\bf K}$ always mean
three-dimensional vectors in unit of $(2\pi/a,2\pi/a,2\pi/c)$,
while ordering vectors ${\bf q}$ or ${\bf Q}$ are two-dimensional
vectors in unit of $(2\pi/a,2\pi/a)$. 
To discuss properties in
the presence of external magnetic field like magnetic susceptibility or temperature--magnetic field
phase diagram, we introduce an interlayer interaction only in the present case as follows:
\begin{eqnarray}
{\mathcal H}_{c} = \frac 12 \Lambda_z^\perp \sum_{l}
J_{z,i}^{(l)} J_{z,i}^{(l+1)},
\end{eqnarray}
where we have explicitly introduced the layer index
$l$ in this case, and the factor $1/2$ accounts for the number of nearest neighbors along the $c$ axis.
Although the Ising-type interaction has been chosen for simplicity, there is no difference of the result even if we choose an isotropic interlayer interaction. 
The Ising-type AFM transition temperature $T_z$ is given by
\begin{equation}
1-(\Lambda_z+\Lambda_z^{\perp}) \chi_L(T_z)= 0,
\label{eq:tz}
\end{equation}
where we have assumed $\Lambda_z^{\perp} >0$, and
the local susceptibility $\chi_L(T)$ is given by
$\chi_L(T)= \beta
[1+{\rm exp}(-\beta\Delta)/2]^{-1}$
with $\beta=1/T$.
The (homogeneous) magnetic susceptibility for $H||(001)$ in the paramagnetic phase ($T>T_z$) is given by
\begin{eqnarray}
\chi (T) = \frac{
\chi_L(T)} {
1-(\Lambda_z-\Lambda_z^{\perp}) \chi_L(T) }\,.\label{eq:susc}
\end{eqnarray}
Note that the antiferromagnetic interplane interaction $\Lambda_z^{\perp} >0$ reduces the homogeneous susceptibility.

\begin{figure}[t]
\centering
\includegraphics[totalheight=6.5cm,angle=270]{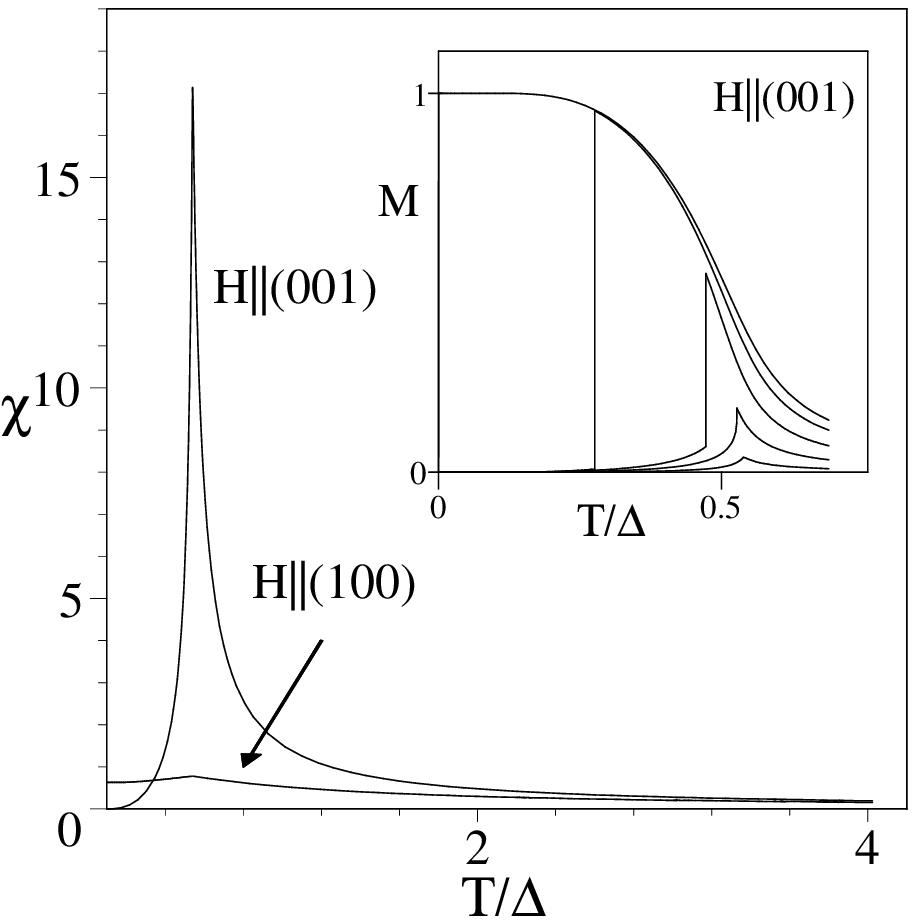}
\includegraphics[totalheight=6.5cm,angle=270]{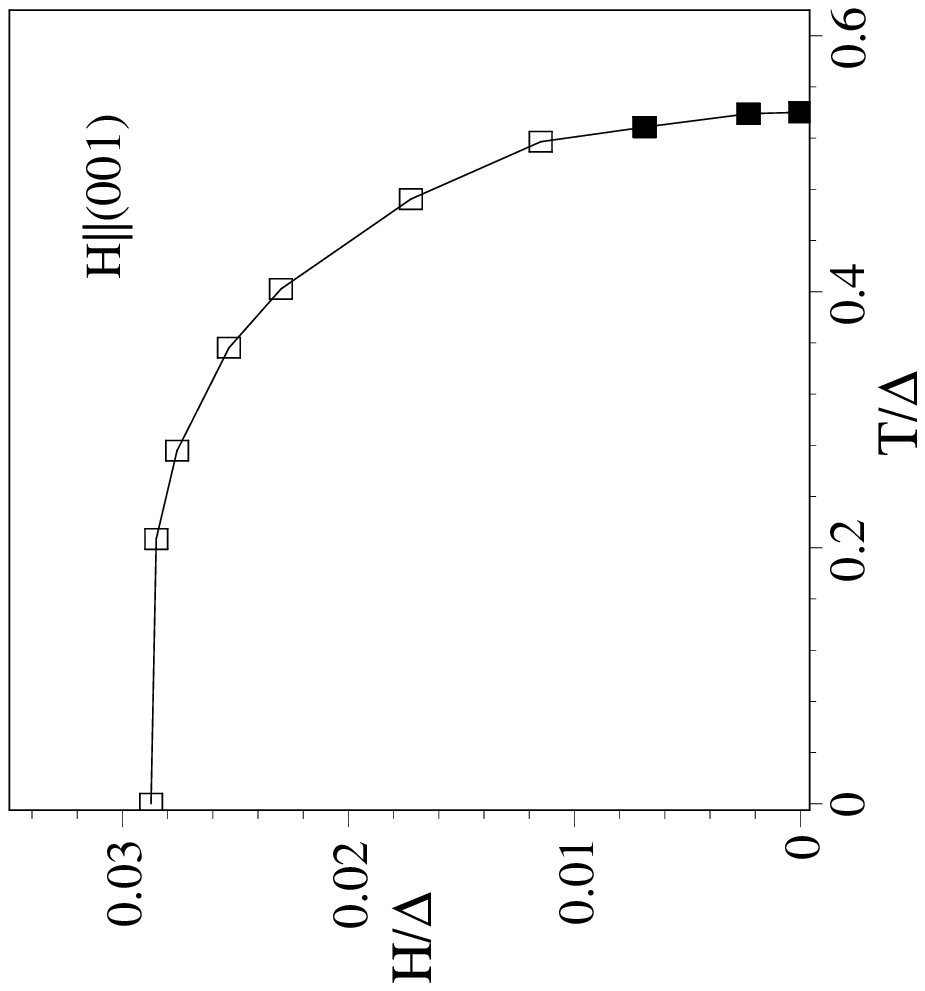}
\caption{{\sl Left:} Temperature dependence of the magnetic susceptibility in the Ising doublet limit.
{\sl Inset} shows the magnetization curves as a function of temperature for
fields $h=0.002$, 0.008, 0.017, 0.027 and 0.034. {\sl Right:} $T$--$H$ phase diagram for field direction $(001)$.
Filled squares mean second order transitions, while empty squares represent first order one. We took
$\lambda_z+\lambda_z^{\perp}=0.586$ with $\lambda_z^{\perp}/\lambda_z=0.05$ in the calculation, which gives the zero field phase transition as $T_z/\Delta =0.54$.}\label{fig:1}
\end{figure}

The left part of Fig.~\ref{fig:1} shows the magnetic susceptibility for both magnetic field directions (001) and (100). We use dimensionless coupling constants and magnetic field such as
 $\lambda_{k,\mu}\equiv \Lambda_{k,\mu}/\Delta$,
and $h\equiv
g\mu_{\rm B}H/(k_{\rm B}\Delta)$, where $g=3/5$.
If we decrease the ratio $\lambda_z^{\perp}/\lambda_z$, i. e., towards weaker interlayer coupling, the peak in the magnetic susceptibility for $H||(001)$ becomes sharper because of the decrease of the denominator in expression (\ref{eq:susc}).

We also calculated the temperature dependence
of the magnetization, which shows metamagnetic transition at high
fields and low temperatures as shown in the inset of Fig.~\ref{fig:1}. The calculated temperature--magnetic field phase diagram can be seen in the right part of Fig.~\ref{fig:1}.
Let us consider the metamagnetic transition in the ground state. The staggered component of the order parameter $\langle J_z\rangle_A-\langle J_z\rangle_B$ vanishes when the magnetic field exceeds a critical value $H_{cr}$.
For $H<H_{cr}$ the ground state energy in the mean-field theory can be written as
\begin{eqnarray}
{\cal F}_{<} =
\frac{1}{4}\lambda_z\left(\langle J_z\rangle_A^2+\langle J_z\rangle_B^2\right)
- \frac{1}{2}\lambda_z^{\perp}\langle J_z\rangle_A\langle J_z\rangle_B-
\frac{1}{2}
(\lambda_z+\lambda_z^{\perp})
\left(\langle J_z\rangle_A-\langle J_z\rangle_B\right)
,\label{eq:afm1}
\end{eqnarray}
while for $H>H_{cr}$ it becomes
\begin{eqnarray}
{\cal F}_{>} = \frac{1}{2}
(\lambda_z-\lambda_z^{\perp})\langle J_z\rangle^2-(\lambda_z-\lambda_z^{\perp})\langle J_z\rangle -H.
\label{eq:afm2}
\end{eqnarray}
Within the AFM ordered phase the conditions $\partial {\cal F}_{<}/ \partial \langle J_z\rangle_A=0$ and $\partial {\cal F}_{<}/ \partial \langle J_z\rangle_B=0$ gives $\langle J_z\rangle_A=-\langle J_z\rangle_B=1$, which leads to ${\cal F}_{<} =-1/2(\lambda_z+\lambda_z^{\perp})$.
Similarly, in the high-field phase we get $\langle J_z\rangle=1$ from (\ref{eq:afm2}), which gives ${\cal F}_{>} =-1/2(\lambda_z-\lambda_z^{\perp})-H$.
The AFM order vanishes discontinuously by increasing the magnetic field at $H=H_{cr}$, which gives ${\cal F}_{<}= {\cal F}_{>}$.
From expressions (\ref{eq:afm1}) and (\ref{eq:afm2}) with $\langle J_z\rangle=1$ we get $H_{cr}=\lambda_z^{\perp}$. Namely $H_{cr}$ is given by the value of the interlayer coupling constant.
This relation clarifies the origin of the metamagnetic transition.

Expression (\ref{eq:tz}) shows that the transition temperature  $T_z$ is related to the value of $\lambda_z+\lambda_z^{\perp}$, while the critical magnetic field $H_{\rm cr}$ to $\lambda_z^{\perp}$.
Therefore, $H_{\rm cr}$ can be small relative to the transition temperature $T_z$ if we choose small interlayer coupling $\lambda_z^{\perp}$.
Furthermore, with small $\lambda_z^{\perp}$, the susceptibility is sharply peaked  for $H||(001)$. These features are in good agreement with the measured results for NpCoGa$_5$\cite{aoki,metoki, colineau}.

\subsection{Case~II}

In this limit we take nonzero dipole interactions
$\Lambda_z$ and $\Lambda_x =\Lambda_y$, and
assume that the quardupolar interaction $\Lambda_{xy}$ dominates over
$\Lambda_{yz}=\Lambda_{zx}$.
This $\Lambda_{xy}$ quadrupolar interaction stabilizes the magnetic moment along $(110)$
because of the third order term
$\langle J_x\rangle \langle J_y\rangle \langle O_{xy}\rangle$
in the Landau free-energy expansion.
Namely,  the system gains the maximum energy when both $\langle J_x\rangle$ and $\langle J_y\rangle$ are non-zero and equal.
The third order term also shows that pure magnetic order of the perpendicular dipoles $\langle J_x+J_y
\rangle$ does not exist,
since the quadrupoles $\langle
{O}_{xy}\rangle$ are induced by the symmetry.
\begin{figure}[ht]
\centering
\includegraphics[totalheight= 6cm,angle=0]{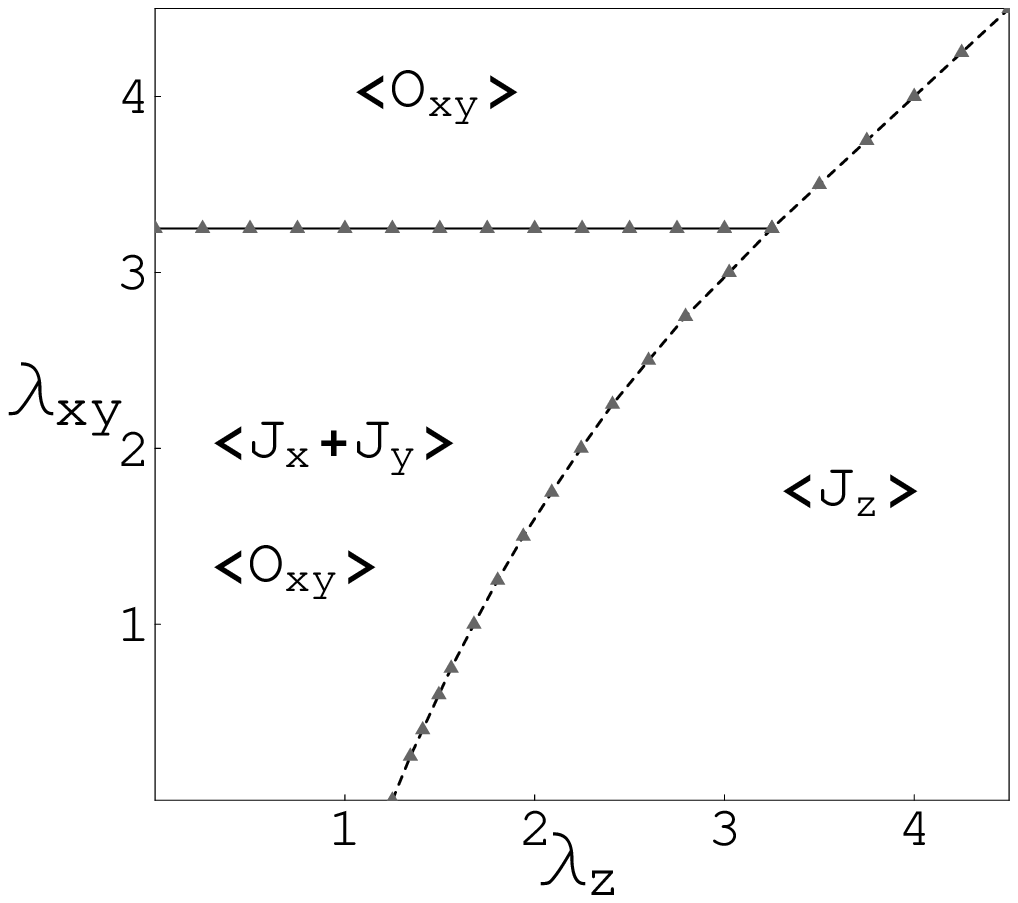}
\includegraphics[totalheight= 6cm,angle=0]{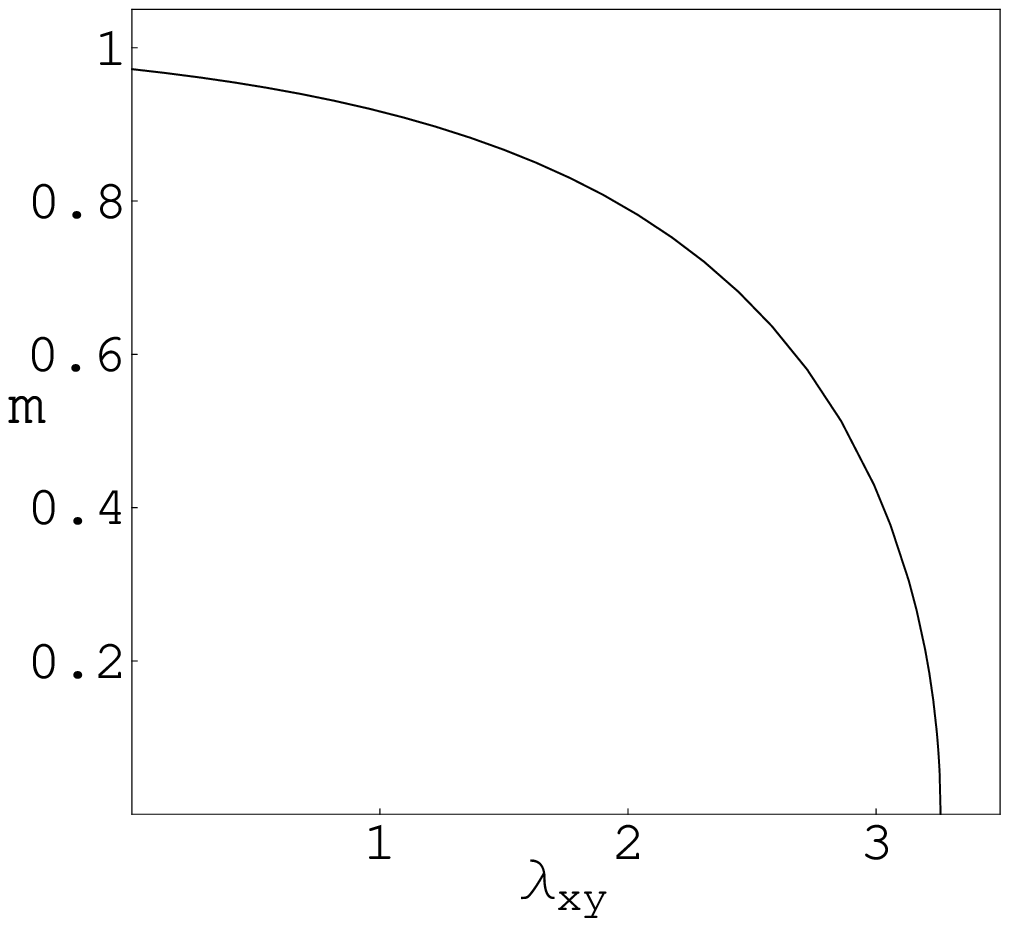}
\caption{{\sl Left:} The ground-state phase diagram with
$\lambda_x=2.13$.  For details see the text. {\sl Right:} Magnetic moment $m=\langle J_x+J_y \rangle/\sqrt{2}$ in the ground state as a function of $\lambda_{xy}$ for $\lambda_x=2.13$ and $\lambda_z<1.25$.}\label{fig:2}
\end{figure}

The left part of Fig.~\ref{fig:2}
shows the ground state of the model
with fixed $\lambda_{x} =\lambda_{y} =2.13$, which value is chosen as a trial,
and $\lambda_{zx} =\lambda_{yz} =0$.
In the limit of $\lambda_{xy}, \lambda_z \gg \lambda_x$,
the phase boundary between the magnetic ($\langle J_z\rangle\neq 0$) and the quadrupolar ($\langle O_{xy}\rangle\neq 0$) phases tends to
$\lambda_{z}=\lambda_{xy}$, recovering the doublet limit with $\lambda_x=0$.
The nonzero $\lambda_x$
causes the deviation of the
phase boundary from $\lambda_{z}=\lambda_{xy}$ in such a way that the mixed phase with $\langle J_x+J_y \rangle$$\langle O_{xy}\rangle\neq 0 $ expands for $\lambda_{xy} \lesssim \lambda_{x}$ at the expense of phase with $\langle J_z\rangle\ne 0$ in the ground state.
The right part of Fig.~\ref{fig:2} shows the ground state magnetic moment
\begin{eqnarray}
m=\frac{1}{\sqrt{2}} \langle J_x+J_y \rangle = 2 \left[ \frac{4\lambda_{x}^2-1-\lambda_{xy}-2\lambda_{x}\lambda_{xy}}{16\lambda_{x}^2-8\lambda_{x}\lambda_{xy}+\lambda_{xy}^2} \right]^{1/2}\label{rh_mom}
\end{eqnarray}
as a function of the quadrupolar interaction $\lambda_{xy}$ within the mixed phase.
We can see that the ordered magnetic moment decreases with increasing quadrupolar interaction, and finally disappears at $\lambda_{xy}=3.26$ in the case of $\lambda_{x} =2.13$. The vanishing of the moment gives the second order phase boundary between the pure quadrupolar phase and the mixed phase. The boundary is given by $\lambda_{xy}=2\lambda_x-1$, which can be seen as the horizontal straight line on the phase diagram shown in the left part of Fig.~\ref{fig:2}.

\begin{figure}[t]
\centering
\includegraphics[height=6.5cm]{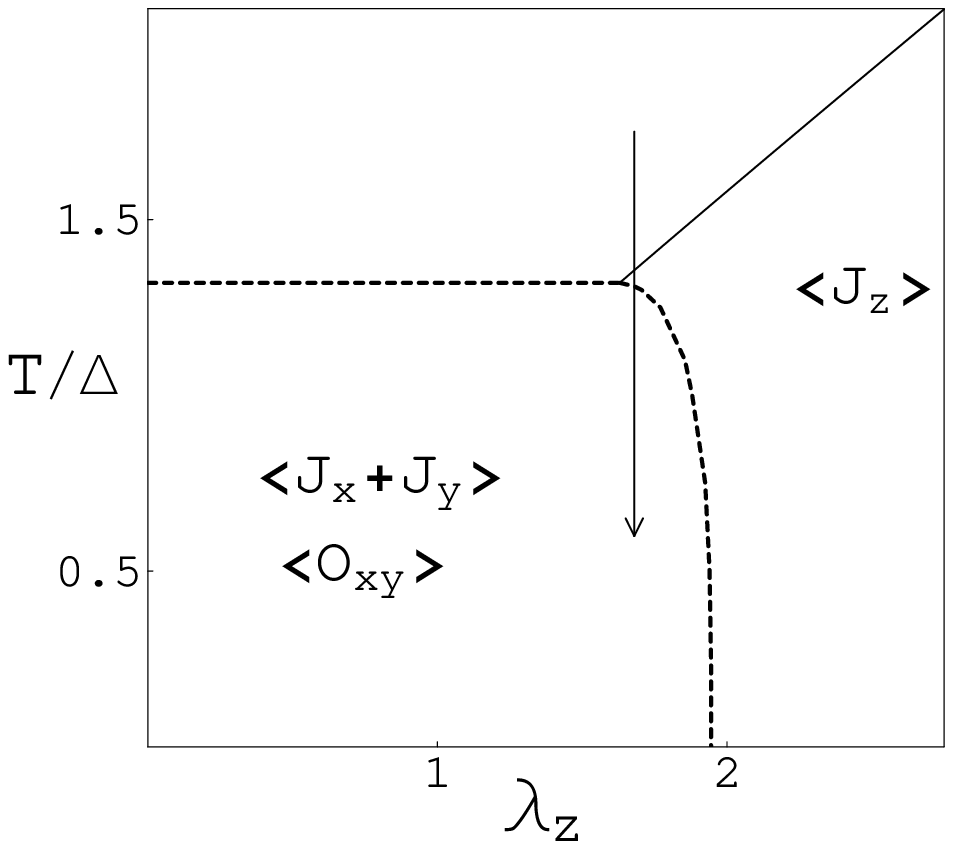}
\includegraphics[height=6.5cm]{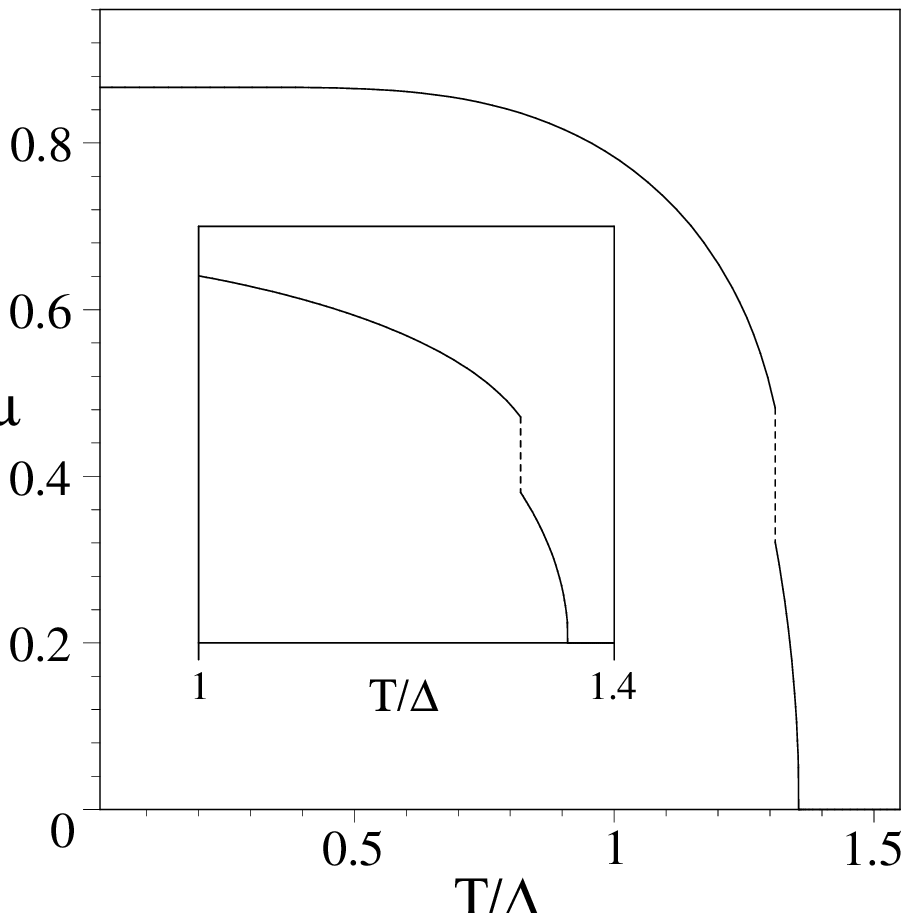}
\caption{
{\sl Left:}
$\lambda_z$--$T$ phase diagram for
$\lambda_{xy}=1.5$ and $\lambda_x=2.13$.
Dashed lines indicate the first order transitions, while the
solid line represents the second order one.
Their meeting point represents a multicritical point which is near the arrow relevant to NpRhGa$_5$.
{\sl Right:}
Temperature dependence of the magnetic moment along the arrow in the upper panel.
With $\lambda_z+\lambda_z^{\perp}=1.68$, the phase transitions are derived as $T_z/\Delta=1.31$ and $T_x/\Delta=1.36$. {\sl Inset} shows the moment near the transitions.}\label{fig:3}
\end{figure}

The compound NpRhGa$_5$ shows two
successive magnetic transitions.
From neutron diffraction it is concluded
that the magnetic structure below the first transtion
$T_{\rm N}^1=36$K is the same
as that in NpCoGa$_5$, while
at $T_{\rm N}^2=33$K the moment
changes the direction to $\langle J_x+J_y\rangle\ne 0$ by a first order transition.
The moments are ferromagnetically ordered within each $ab$--plane\cite{jonen,aoki2,colineau2}.
The ordered moments $\langle J_z\rangle$ completely vanish below 33K.
In order to simulate the situation in NpRhGa$_5$ qualitatively, we now fix a part of parameters as $\lambda_{xy}=1.5$ and $\lambda_{x}=2.13$, and derive the phase diagram in the $\lambda_z$--$T$ plane.
In the parameter range $\lambda_{xy} \lesssim \lambda_{x}$ the quadrupolar interaction prefers
in-plane magnetic moments to the $c$-axis moments.
A consequence is that there appears a regime on the $\lambda_z$--$T$
phase diagram where
an ordered phase with
$\langle J_z\rangle \ne 0$ undergoes a first order transition
 to another one with $\langle J_x+J_y \rangle \langle O_{xy} \rangle\ne 0$
with decreasing temperature. The left part of Fig.~\ref{fig:3} shows the calculated phase diagram. Along the path ($\lambda_z=1.68$) shown by the arrow,
the moment $\mu=\sqrt{(J_x+J_y)^2/2+J_z^2}$ develops as shown in
the right panel.
The inter-plane AFM coupling $\lambda_z^{\perp}$
can simply be included in the mean-field theory
by modifying
$\lambda_z \rightarrow \lambda_z +\lambda_z^{\perp}$.
The magnetic moment increases discontinuously at the first order transition.
This behavior is in qualitative agreement with the results found
in NpRhGa$_5$ \cite{jonen}.
Note that with $\lambda_z +\lambda_z^{\perp} \sim 1.62$,  the three phases meet at the multicritical point.

We note that the consideration of the singlet state is essential to obtain the perpendicular magnetic order in Case II. This is in contrast to the case of NpCoGa$_5$, which can be described in the doublet limit.
However, the value of the energy separation $\Delta$ is not crucial to the existence of the phases obtained and their characteristic features.

Although we considered a two-dimensional model taking into account
only the interactions within the $ab$-planes, we comment shortly on the
three-dimensional magnetic and quadrupolar structure. The magnetic
moments $\langle J_x+J_y\rangle$ are ferromagnetically ordered
within the $ab$ planes at low temperatures. This means that
homogenous quadrupolar moments $\langle O_{xy}\rangle$ are present
within the planes even when the quadrupolar coupling constant
$\lambda_{xy}$ is zero. The interplane stacking of the magnetic
moments is antiferromagnetic, but the ferroquadrupolar order is
realized on subsequent planes. This situation is consistent with
the invariant term $\langle J_x\rangle({\bf k})\langle
J_y\rangle({\bf -k})\langle O_{xy}\rangle({\bf 0})$ in the Landau
free energy expansion. The ferroquadrupolar order means that an
orthorombic lattice distortion
should develop below $T_{\rm N}^2=33$K.
We expect experimental observation of the distortion.

\subsection{Case~III}

Now we assume that the quadrupolar interactions
$\Lambda_{zx} =\Lambda_{yz}$ dominate over $\Lambda_{xy}$ in contrast with cases of T=Co or Rh.
In this case the third order term
$\langle J_z \rangle
(\langle J_x\rangle \langle O_{zx}\rangle+\langle J_y\rangle \langle O_{yz}\rangle)$
in the Landau expansion plays an important role. Namely, it can lead to two successive transitions both of which are continuous.
After the ordering of one of the multipole moments $J_x$, $J_z$ or $O_{zx}$, a lower phase transition is possible by the simultaneous appearance of the other two multipole moments.
A difficulty of the ordinary Landau-type expansion is that the order parameter of the first phase transition is not necessarily small at the second transition, so that
the expansion of the free energy with respect to it is not justified.
Therefore we use the method developed in ref.\citen{kuramoto1},
and calculate the Ginzburg-Landau free energy functional in all orders of the finite order parameter which becomes nonzero below the first phase transition.
In the present limit we consider the following interactions within the tetragonal $ab$--plane:
\begin{eqnarray}
{\mathcal H} =
-\frac 14\sum_{\langle ij\rangle} \left( \Lambda_z J_{z,i}J_{z,j}+\Lambda_x J_{x,i}J_{x,j}+\Lambda_{zx} O_{zx,i} O_{zx,j} \right) \,.
\label{eq:ham103}
\end{eqnarray}
We start with the path-integral representation of the partition
function\cite{kuramoto1}
\begin{eqnarray}
{\cal Z}=\int {\cal D}X^A {\rm exp}\left[-S_{\rm B}
-\int_{0}^{\beta} d\tau {\cal H}(\tau)\right]\,,
\end{eqnarray}
where the integration variable $X^A$ replaces the operators $J_x$, $J_z$ and $O_{zx}$.
$S_{\rm B}$ is the Berry phase term, 
which enters because of the non-bosonic commutation property of $J_x$, $J_z$ and $O_{zx}$.
We use the Hubbard-Stratonovich transformation for
each imaginary time interval, which is the replacement $-\Lambda_A
X^A\rightarrow \phi^A$ by introducing effective fields $\phi^A$.
These fields $\phi^A$ mediate the original interaction. The
Hubbard-Stratonovich transformation converts the original
inter-site interaction into a local interaction between the
effective fields and multipolar operators. Using the property that
$\phi$'s have Gaussian distribution, the partition function can be
expressed as
\begin{eqnarray}
{\cal Z}=\int {\cal D}X^A{\cal D}\phi^A {\rm exp}\left[ -S_{\rm
B}-\int_{0}^{\beta} d\tau {\cal H}_{\phi}(\tau)\right]\,,
\end{eqnarray}
where
\begin{eqnarray}
{\cal H}_{\phi}=\frac{1}{4}\sum_{i, j} \Lambda_{A} \phi_i^A \phi_j^A-\frac{1}{2}\sum_{i}\phi_i^A X_i^A\,.
\end{eqnarray}
We further use the static approximation within which the path integral over $X^A$ in ${\cal Z}$ is replaced by trace calculation of corresponding operators\cite{kuramoto3}.
Then we obtain the following form for the partition function
\begin{eqnarray}
{\cal Z}=\int {\cal D}\phi^A {\rm exp}[-\beta {\cal F}]\,,
\end{eqnarray}
 where ${\cal F}$ is the Ginzburg-Landau free energy functional given by
 \begin{eqnarray}
 {\cal F} &=& -N\frac{1}{\beta}{\rm ln}{\cal Z}_0+ Nf_{\rm mc}\nonumber\\
 &+& \frac{1}{4}{\sum_{i,j}}\delta_{ij} \left[\left(\Lambda_{x}^{-1} +d_1 \right) \phi_i^x \phi_j^x
 + \left(\Lambda_{z}^{-1}  +d_2\right)\phi_i^z \phi_j^z +
 \left(\Lambda_{zx}^{-1} +d_1 \right)\phi_i^{zx} \phi_j^{zx}\right] \label{eq:free55}
 \end{eqnarray}
with
\begin{eqnarray}
d_1=\frac{2}{\Delta}\frac{({\rm e}^{-\beta \Delta}-1)}{(2+{\rm e}^{-\beta \Delta})},\hspace*{0.3cm}
d_2=-\beta \frac{2}{(2+{\rm e}^{-\beta \Delta})}\,,\nonumber
\end{eqnarray}
and ${\cal Z}_0={\rm exp}(-\beta \Delta)+2$. $N$ is the number of the sites.
The mode coupling free energy has the form
\begin{eqnarray}
 f_{\rm mc} = -\frac{1}{N}\frac{1}{\beta}\sum_{i} \left[ {\rm ln}\left( {\rm Tr}_i {\rm exp} (\beta \phi_i^A X_i^A/2)/{\cal Z}_0\right) +\frac{1}{4}\beta d_1\phi_i^x \phi_i^x
 +\frac{1}{4}\beta d_1\phi_i^{zx} \phi_i^{zx} +\frac{1}{4}\beta d_2 \phi_i^z \phi_i^z \right]
 \,.\label{eq:mc14}
 \end{eqnarray}

Let us consider the case where the first phase transition at a
temperature $T=T_{\rm H}$ corresponds to the ordering of $J_z$
dipole moments. This means that for $T<T_{\rm H}$ the expectation
value $\langle \phi^z_{\bf Q} \rangle=\Lambda_z({\bf Q})\langle
J_z \rangle({\bf Q})$ becomes nonzero, where $\Lambda_z({\bf
Q})=(1/2)\Lambda_z({\rm cos}(Q_x)+{\rm cos}(Q_y))$ is the Fourier
transform of the coupling constant $\Lambda_z$. The lower
transition temperature $T_{\rm L}$ is derived from the condition
${\rm det}(\hat{\chi}^{-1})=0$, where $\hat{\chi}$ is the
generalized susceptibility matrix.
We have the relation:
\begin{eqnarray}
\hat{\chi}= \hat{J}^{-1} \hat{G} \hat{J}^{-1} - \hat{J}^{-1}\,,\label{eq:gensusc}
\end{eqnarray}
where the interaction matrix $\hat{J}$ is given by
\begin{eqnarray}
\hat{J}=  \left( \begin{array}{ccc}
       \Lambda_{z}({\bf Q}) & 0 & 0\\
       0 &  \Lambda_{x}({\bf q})  &  0 \\
           0 & 0 & \Lambda_{zx}(-({\bf q+Q})) \end{array}
       \right)\,.
\label{eq:matrix}
\end{eqnarray}
The matrix $\hat{G}$ satisfies the
following equation~\cite{kuramoto2}
\begin{eqnarray}
\hat{G}= \beta \left( \begin{array}{ccc}
        \langle \phi_{\bf Q}^z \phi_{-{\bf Q}}^z\rangle & 0 & 0\\
        0 &  \langle \phi_{\bf q}^x \phi_{-{\bf q}}^x\rangle   & \langle \phi_{\bf q}^x \phi_{-({\bf q+Q})}^{zx}\rangle \\
          0 &  \langle \phi_{-({\bf q+Q})}^{zx} \phi_{\bf q}^x\rangle & \langle \phi_{{\bf q+Q}}^{zx} \phi_{-({\bf q+Q})}^{zx} \rangle \end{array}
       \right)\,,
\label{eq:matrix}
\end{eqnarray}
which means the generalized susceptibility for the effective fields.
We are interested in the case where the first transition is ferro-type
with ${\bf Q}=0$. Therefore, the wave vector ${\bf q}$ is equivalent to ${\bf q+Q}$.

If the lower transition is also of second order, we make 
perturbation expansion,
under the finite value of $\langle \phi^z_{0} \rangle$, 
 for the mode coupling free energy (\ref{eq:mc14}) in terms of order parameters $\phi^x_{\bf q}$ and $\phi^{zx}_{\bf q}$ up to second order\cite{kuramoto1}.
Then, the Ginzburg-Landau free energy functional can be written in the form
\begin{eqnarray}
{\cal F}=\frac{1}{4} H_{11}\phi^z_0 \phi^z_0 +  \frac{1}{4} \sum_{i,j=2}^3 H_{ij} \phi^i_{\bf q} \phi^j_{-{\bf q}}\,,\label{eq:freeh}
\end{eqnarray}
where the matrix $\hat{H}=\{ H_{ij} \}$ composed by the coefficients of the second order terms is related to $\hat{G}$ as $\hat{H}^{-1}=\hat{G}$, and it has the same block diagonal form.
We note when $\phi^i$ shows Gaussian distribution, $G^{ij}$ is given by its variance. Using the relation (\ref{eq:gensusc}) 
we obtain
\begin{eqnarray}
{\rm det}(\hat{\chi}^{-1})=\frac{\Lambda_{z}(0)\Lambda_{x}({\bf q})\Lambda_{zx}({\bf q})  H_{11}(H_{22}H_{33}-H_{23}H_{32})}{[\Lambda_z(0)^{-1}-H_{11}]
[(\Lambda_x({\bf q})^{-1}-H_{22})(\Lambda_{zx}({\bf q})^{-1}-H_{33})-H_{23}H_{32}]}\,.\label{eq:detchi}
\end{eqnarray}
Therefore, the condition ${\rm det}(\hat{\chi}^{-1})=0$ is equivalent with ${\rm det}(\hat{H})=0$. Calculating the coefficients $H_{ij}$ in the free energy expansion (\ref{eq:freeh}), 
we can derive the lower phase transition by the condition ${\rm det}(\hat{H})=0$.

\begin{figure}
\centering
\includegraphics[totalheight=5.9cm,angle=0]{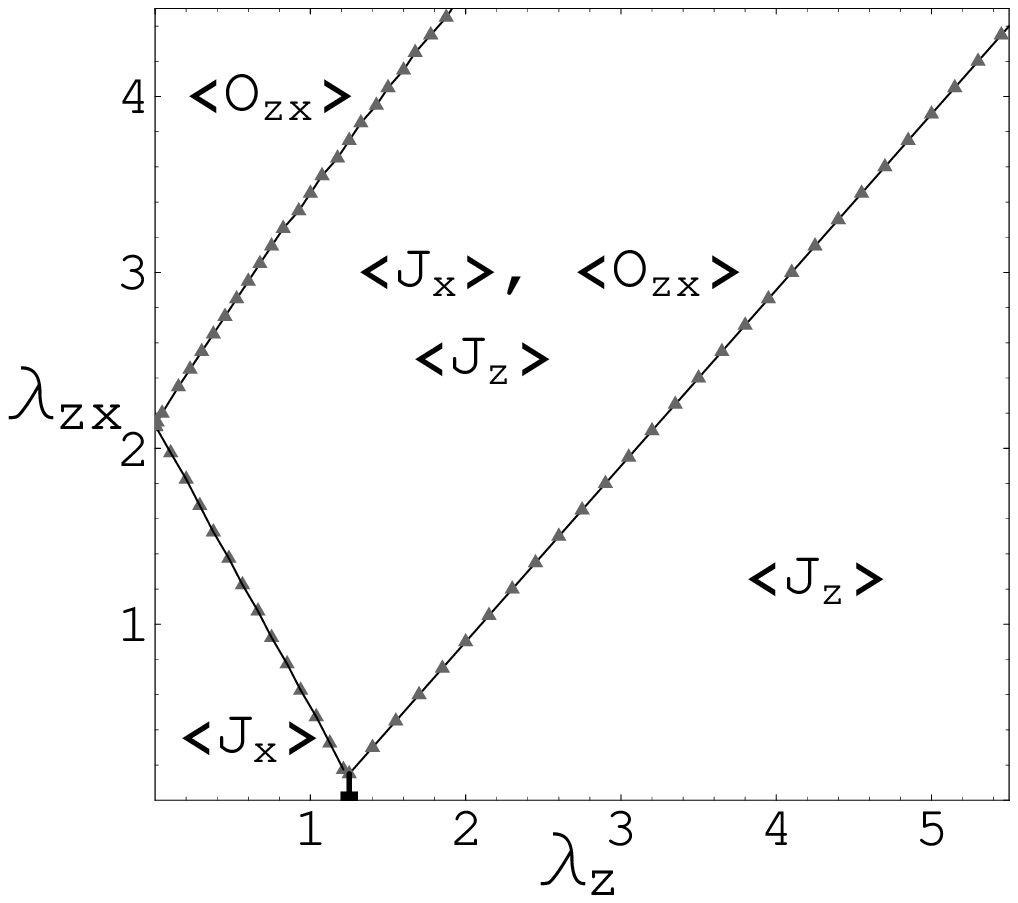}
\includegraphics[totalheight=6.4cm,angle=0]{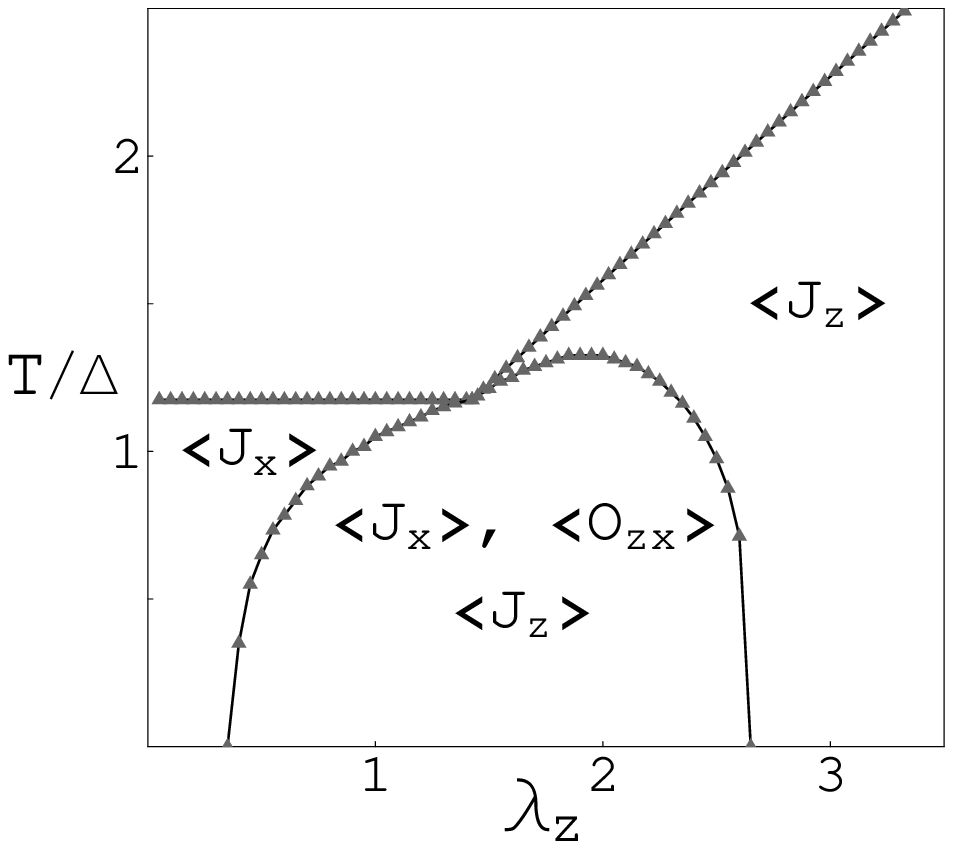}
\caption{{\sl Left:}  Finite order parameters at $T=0$
in the $\lambda_z$--$\lambda_{zx}$ plane with $\lambda_x=2.13$.
{\sl Right:}  Phase diagram in the $\lambda_z$--$T$ plane with a tetracritical point.
We have chosen $\lambda_x=2.13$ and $\lambda_{zx}=1.5$.}\label{fig:4}
\end{figure}

\begin{itemize}
\item $T \geq T_{\rm H}$\\
In this temperature regime with $\langle \phi^z_{0}\rangle=0$, 
the mode coupling free energy (\ref{eq:mc14}) gives $H_{23}=H_{32}=0$ in (\ref{eq:freeh}).
The second order coefficient of term $\phi^z_{0}$ can be read from (\ref{eq:free55}) as
\begin{eqnarray}
H_{11}=\frac{1}{\Lambda_z}-\beta\frac{2}{(2+{\rm e}^{-\beta \Delta})}=\frac{1}{\Lambda_z}-\chi_{L}\,,
\end{eqnarray}
where we have used that $\Lambda_z(0)=\Lambda_z$.
Due to the block form of matrices $\hat{G}$ and $\hat{H}$, the component of the generalized susceptibility, which corresponds to the dipole operator $J_z$ can be obtained as
\begin{eqnarray}
\chi_{11}=\Lambda_z^{-1}(H_{11}^{-1}-1)=
\frac{\chi_L} {
1-\Lambda_z \chi_L }\,,
\end{eqnarray}
which is the same as expression (\ref{eq:susc}) derived in Case~I with $\Lambda_z^{\perp}=0$.
This is due to the fact that
the homogenous magnetic field lying in direction $(001)$ is the conjugated field to the dipole moment $J_z$ in the case of ${\bf K}=0$.
The first transition temperature $T_{\rm H}$ can be derived from $H_{11}=0$,
as the first instability in ordinary Landau mean field theory 
(see equation (\ref{eq:detchi})). 
This leads to the condition $1-\Lambda_z\chi_{L}(T_{\rm H})=0$, which is the same as expression (\ref{eq:tz}) taking $\Lambda_z^{\perp}=0$. Or equivalently, $\chi_{11}^{-1}=0$ leads to the same condition for the transition temperature.

\item $T_{\rm H} >T \geq T_{\rm L}$\\
In this temperature regime $f_{\rm mc}$ becomes nonzero in the Ginzburg-Landau free energy functional (\ref{eq:free55}).
Under the finite value of $\langle \phi^z_{0} \rangle=\Lambda_z \langle J_z\rangle$, the Ginzburg-Landau free energy functional is derived as
\begin{eqnarray}
{\cal F}  &=& \frac{1}{4}\sum_{\bf q}\left[ (\Lambda_x({\bf q})^{-1}-\pi_1)\phi_{{\bf q}}^{x}\cdot \phi_{-{\bf q}}^{x}+(\Lambda_{zx}({\bf q})^{-1}-\pi_1)\phi_{{\bf q}}^{zx}\cdot \phi_{-{\bf q}}^{zx}
-\pi_2\phi_{{\bf q}}^{x}\cdot \phi_{-{\bf q}}^{zx} \right. \nonumber\\
&&\left.+(\Lambda_z^{-1}+d_2)\phi^z_0\phi^z_0-d_2\langle \phi^z_0\rangle^2 \right] \,,\label{eq:free97}
\end{eqnarray}
where
\begin{eqnarray}
\pi_1 &=&\frac{1}{{\tilde {\cal Z}_0}}\left(  \frac{{\rm e}^{\beta  \langle \phi^z_{0} \rangle}}{\langle \phi^z_{0} \rangle+\Delta} - \frac{{\rm e}^{-\beta \langle \phi^z_{0} \rangle}}{\langle \phi^z_{0} \rangle-\Delta}+2\Delta\frac{{\rm e}^{-\beta \Delta}}{\langle \phi^z_{0} \rangle^2-\Delta^2}\right)\,,\nonumber\\
\pi_2 &=&\frac{1}{{\tilde {\cal Z}_0}}\left( \frac{{\rm e}^{\beta \langle \phi^z_{0} \rangle}}{\langle \phi^z_{0} \rangle+\Delta} + \frac{{\rm e}^{-\beta \langle \phi^z_{0} \rangle}}{\langle \phi^z_{0} \rangle-\Delta}-2 \langle \phi^z_{0} \rangle\frac{{\rm e}^{-\beta \Delta}}{\langle \phi^z_{0} \rangle^2-\Delta^2}\right)\,,
\end{eqnarray}
and ${\tilde {\cal Z}_0}={\rm exp}(-\beta \Delta)+2{\rm cosh} ( \beta \langle \phi^z_{0} \rangle )$.
When $\langle \phi^z_{0} \rangle \rightarrow 0$ 
we obtain $\pi_2\rightarrow 0$ and $\pi_1=-d_1$, which means the vanishing of the mixing term as we expect.
We then come back to the high temperature form of the GL free energy functional given by (\ref{eq:free55}) with $f_{\rm mc}=0$.
The lower transition temperature $T_{\rm L}$ can be derived from ${\rm det}(\hat{\chi}^{-1})=0$ or ${\rm det}(\hat{H})=0$.
Explicitly we have the condition
\begin{eqnarray}
[\Lambda_x({\bf q})^{-1}-\pi_1(T_{\rm L})][\Lambda_{zx}({\bf q})^{-1}-\pi_1(T_{\rm L})]-\pi_2(T_{\rm L})^2=0\,.\label{eq:condxzx}
\end{eqnarray}
\end{itemize}
Similar calculation can be performed for the free energy expansion under finite $\langle \phi^A_{0} \rangle=\Lambda_x(0) \langle J_x\rangle(0)$ or $\langle \phi^A_{0} \rangle=\Lambda_{zx}(0) \langle O_{zx}\rangle(0)$.
The left panel of Fig.~\ref{fig:4} shows the ground state
($T=0$) phase diagram with $\lambda_x=\Lambda_x/\Delta$ fixed at 2.13.
We find three regimes where pure dipolar ($J_z$ or $J_x$) or quadrupolar ($O_{zx}$) order is realized.
Thus, we can enter to the regime with mixed order parameters
$\langle J_z\rangle$--$\langle J_x\rangle$--$\langle {
O}_{zx}\rangle$ from these three different sides.
Let us consider the limiting cases of the phase diagram.\\
(i) $\lambda_{zx}=0$: there is a first order transition at
$\lambda_z\approx 1.25$ between the two kinds of magnetic phases.
The situation is similar to that in Fig.~\ref{fig:2} with  $\lambda_{xy}=0$.
The phase boundary between the mixed and $\langle J_z\rangle\ne 0$ phases
can be obtained from condition (\ref{eq:condxzx}) by
taking the limit $T\rightarrow 0$, which gives
\begin{eqnarray}
0 &=& \left(  \frac{1}{\Lambda_x} - \frac{1}{\langle \phi^z_{0} \rangle+\Delta}\right)
\left(  \frac{1}{\Lambda_{zx}} - \frac{1}{\langle \phi^z_{0} \rangle+\Delta}\right)-
\left( \frac{1}{\langle \phi^z_{0} \rangle+\Delta}\right)^2\nonumber\\
& =& \frac{1}{\Delta^2}\left( \frac{\lambda_z+1-\lambda_{zx}-\lambda_x}{\lambda_x\lambda_{zx}(\lambda_z+1)}\right)\,,\label{eq:condxzx2}
\end{eqnarray}
where we have used that $\langle \phi^z_{0} \rangle= \Lambda_z=\lambda_z \Delta$ in the ground state. Thus, from (\ref{eq:condxzx2}) the phase boundary is obtained as $\lambda_{zx}=1-\lambda_x+\lambda_z$.
The boundary between the mixed and $\langle J_x\rangle\ne 0$ phases is derived as $\lambda_{zx}=\lambda_x- \lambda_z (1+2\lambda_x)/(2\lambda_x-1)$. The intersection of these two phase boundaries gives the point $\lambda_z\approx 1.25$. \\
(ii)
$\lambda_z=0$: the point
$\lambda_{zx}=2.13 \ (=\lambda_x)$
separates the
$\langle J_x\rangle\ne 0$ dipolar and $\langle O_{zx}\rangle\ne0$ quadrupolar phases.
The phase boundary between the mixed and $\langle O_{zx}\rangle\ne 0$ phases
terminates at
$$
\lambda_{zx}=
\frac12 \lambda_z
+\frac 14 (2\lambda_x+1)+\frac 14 \sqrt{(2\lambda_x-1)^2+4\lambda_z^2+8\lambda_z\lambda_x+12\lambda_z}.
$$

The right panel
of Fig.~\ref{fig:4} shows the phase diagram in
the $\lambda_z$--$T$ plane with fixed parameters
$\lambda_x=2.13$ and $\lambda_{zx}=1.5$.
With $\lambda_z\gtrsim 1.5$,
 the first transition makes $\langle J_z\rangle\neq 0$,
and the second transition gives $\langle J_x\rangle \langle {
O}_{zx}\rangle \neq 0$ continuously.
However
for $\lambda_z< 1.5$, the first transition gives $\langle J_x\rangle\neq 0$, and at the lower transition the term $\langle J_z\rangle \langle {
O}_{zx}\rangle$ becomes nonzero.
The transition temperature of the pure $J_x$ ordering ($\lambda_z< 1.5$) is given by
\begin{eqnarray}
T_x
=-\frac{\Delta}{{\rm ln}(2\lambda_x-2)-{\rm ln}(2\lambda_x+1)},\label{eq:tx}
\end{eqnarray}
which is derived from the vanishing of the coefficient of the second order term $\phi^{x}_i\phi^{x}_j$ in the free energy expansion (\ref{eq:free55}).
The expression (\ref{eq:tx}) leads to the horizontal line $T_x/\Delta\approx 1.8$ as a function of $\lambda_z$ with $\lambda_x=2.13$ (see the phase diagram). The expression (\ref{eq:tx}) also shows that continuous transition to the phase $\langle J_x\rangle\ne 0$ is not possible at any temperature with $\lambda_x<1$, i. e., $\Lambda_x<\Delta$.

The magnetic coupling $\lambda_z$ enhances the lower transition
temperature by the same mechanism as discussed in ref.\citen{libero}.
The phase diagram contains a tetracritical point at $\lambda_z\approx 1.5$, where the four critical lines meet. Phase diagrams with multicritical points should also have experimental interest, because the application of uniaxial pressure or doping can drive the system through a multicritical point to a different regime of the phase diagram with very different physical properties.

We note that the phase diagram presented in the left part of Fig.~\ref{fig:4} remains the same
for the $(110)$-type perpendicular dipoles instead of the $(100)$-type. Namely, there is a continuous degeneracy with respect to the phase transition temperature within the $J_x$--$J_y$ order parameter space due to the tetragonal symmetry. The difference is that the ordered quadrupoles will have nonzero moments $\langle O_{zx}+O_{yz} \rangle$ instead of $\langle O_{zx}\rangle \ne 0$.

In NpNiGa$_5$ there appear two successive continuous phase
transitions at $T_{\rm c}=30$K and $T_{\rm N}=18$K
\cite{honda,aoki2}. The first one is the ferromagnetic ordering of
$J_z$ dipoles with the ordering vector ${\bf K}=(0,0,0)$. At
temperature $T_{\rm N}$ non-zero $\langle J_x\rangle$ component
appears with ${\bf k}=(1/2,1/2,1/2)$, leading to a canted
antiferromagnetic structure. In ref.\citen{honda} two
possibilities are mentioned for the orderings: (i) both
transitions correspond to the Np sublattice or (ii) one of the
transitions related to the Ni sublattice. We assume that both
transitions are caused by the Np ions, which is consistent with
recent neutron diffraction results\cite{metoki}.

\begin{figure}[t]
\centering
\includegraphics[totalheight=6.5cm,angle=270]{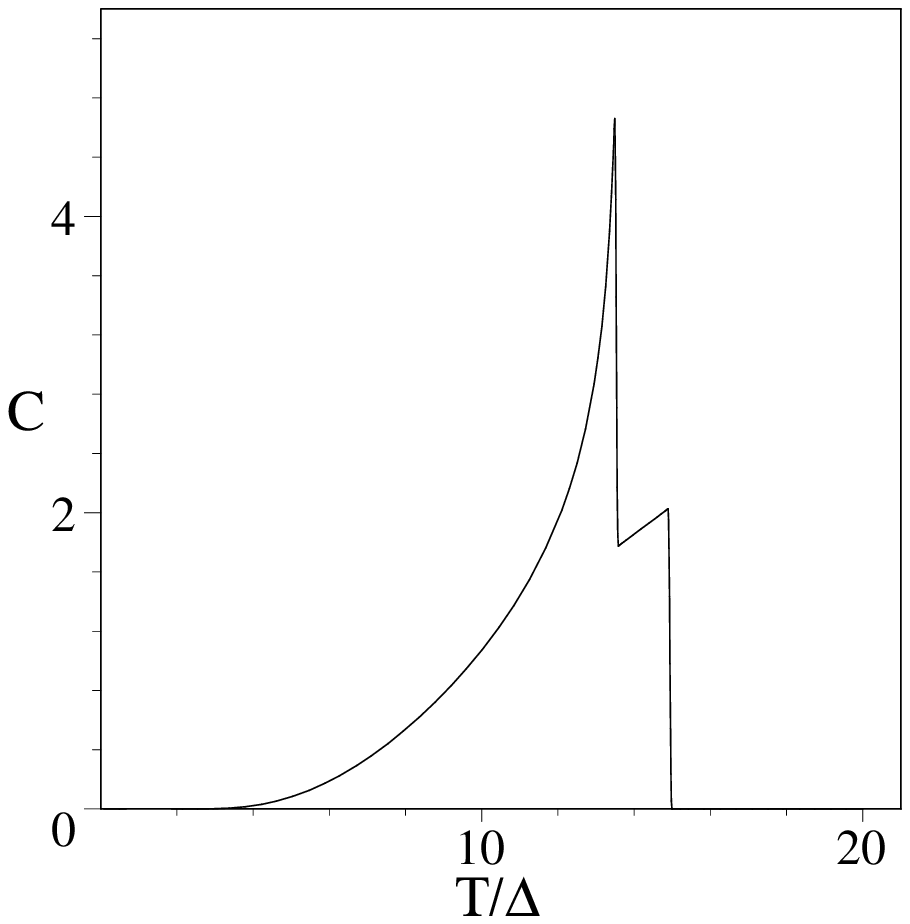}
\includegraphics[totalheight=6.5cm,angle=270]{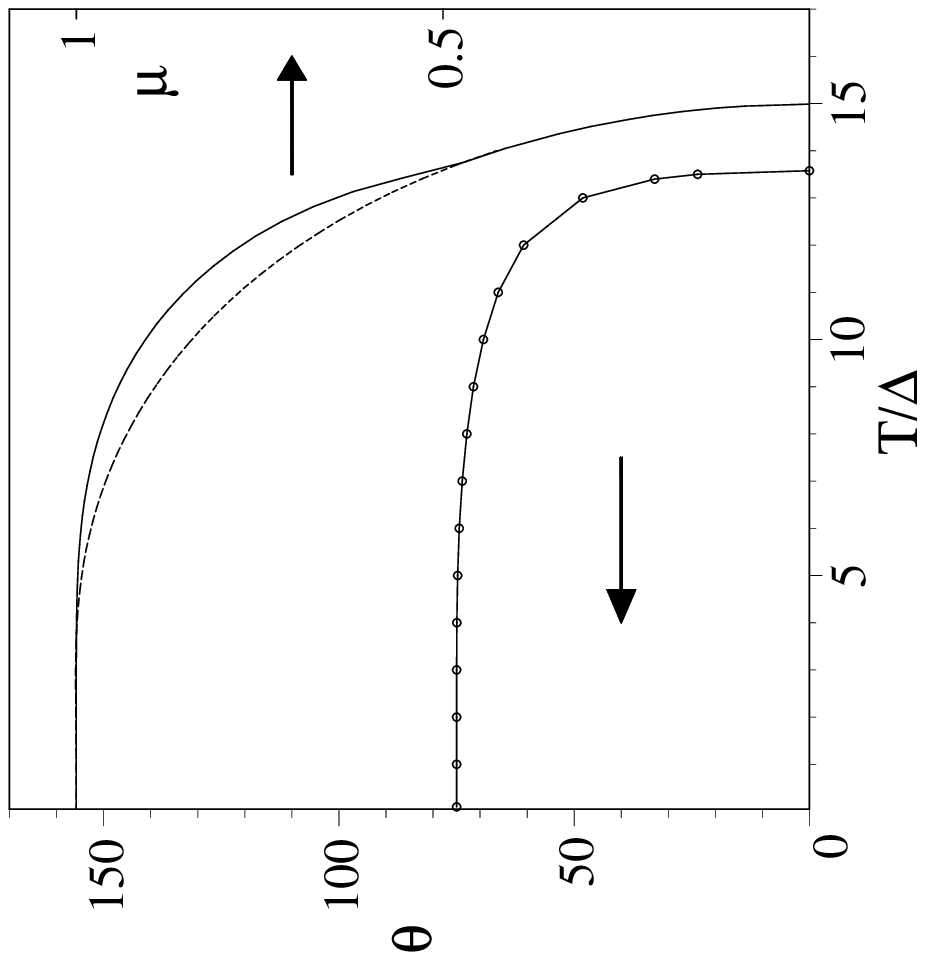}
\caption{
{\sl Left:} Temperature dependence of the specific heat
with two transitions at $T_{z}/\Delta \approx 15$ and $T_{c}/\Delta \approx 13.6$.
The parameters are chosen as
$\lambda_{x}=\lambda_{zx}=16.5$ and $\lambda_{z}=22$.
{\sl Right:} Temperature dependence of $\theta$ (degree) and the moment $\mu$.}\label{fig:5}
\end{figure}

In order to investigate the second transition to
an ordered phase with $\langle J_z\rangle\ne 0$,
we take the simplifying limit $\lambda_x=\lambda_{zx}$. Then the free energy can be calculated by the diagonalization of
the Hamiltonian within the basis (\ref{eq:qt}).
The left panel of Fig.\ref{fig:5} shows the specific heat against temperature with two second-order transitions.
We also calculated the direction of the spontaneous moment ($\theta$) below the lower transition temperature and the total magnetic moment ($\mu$) as a function of temperature (see right part of Fig.\ref{fig:5}).
We define the spontaneous moment
$\overrightarrow{\mu}=\langle J_x\rangle \hat{x}+\langle J_z
\rangle \hat{z}$,
and its deviation from the direction $\hat{z}$
as ${\rm tan}(\theta/2)=\langle J_x\rangle/\langle J_z\rangle$. The
magnitude of the moment is given by
$\mu=\sqrt{\langle J_x\rangle^2+\langle J_z\rangle^2}$.
The calculated temperature
dependences of $\theta$ and $\mu$
are shown in the right panel of Fig.~\ref{fig:5}.
The lower transition causes an anomaly in $\mu$
in good correspondence with the experimental result in NpNiGa$_5$\cite{honda2}.

In the case of antiferromagnetic coupling $\lambda_x<0$, the lower
quadrupolar transition will lead to a canted AFM structure within
the $ab$-planes. In order to have the non-zero invariant term
$\langle J_x\rangle({\bf Q}_{x})\langle O_{zx}\rangle({\bf
Q}_{zx})\langle J_z\rangle({\bf Q}_{z})$ with the ordering wave
vectors in the parenthesis, the condition ${\bf Q}_{zx}=-({\bf
Q}_{x}+{\bf Q}_{z})=-{\bf Q}_{x}$ should be satisfied with ${\bf
Q}_{z}=0$. Hence the quadrupolar order should have the same
ordering vector ${\bf Q}_{x}=(1/2,1/2)$ as that of $J_x$.

\section{Summary and Discussion}

In this work we have studied a two-dimensional mean field model
composed of a non-Kramers doublet ground state and
a singlet excited state. The model includes dipolar and quadrupolar degrees of freedom.
The layered structure of $f$ ions should be important in realizing
characteristic magnetism and superconductivity in $115$ systems.
For theoretical consideration of these systems, a relevant picture
should be a strong two-dimensional interaction and a weak interlayer interaction.
Therefore, the main interactions which determine the magnetic structure are within the two-dimensional $ab$-planes.
The main purpose of this work is to understand the origin of the complex magnetic structures of NpTGa$_5$ systems with T=Co, Ni and Rh. Thus, we have studied a two-dimensional model which includes interactions within the tetragonal $ab$-planes.
We have analyzed some limiting cases of the model by changing the interaction parameters, and found behaviors
reminiscent of the magnetic properties of NpTGa$_5$ systems with T=Co, Ni and Rh.

The intersite interactions in NpTGa$_5$ with T=Ni may be different from those with T=Co or Rh since the Fermi surface of NpNiGa$_5$ contains one more conduction electrons per unit cell. 
The difference in the number of $d$-electrons
may be a reason why
in NpNiGa$_5$ the quadrupolar interactions $\lambda_{zx} =\lambda_{yz}$ dominate over $\lambda_{xy}$ in contrast with cases of T=Co, Rh.
On the other hand,
the CEF structure should be less sensitive to the change of the Fermi surface.
It is likely that other Np 115 systems may correspond to a shifted parameter space.
Then multicritical behaviors can be expected under appropriate conditions. For example, the ordering of the perpendicular magnetic moments can also happen first depending on the interaction parameters. The perpendicular moments experimentally found in NpPtGa$_5$\cite{nppt} seem to be explained in this way.

It is always a question in the case of actinide compounds whether the electrons are localized or itinerant. There is a tendency of increasing localization with increasing atomic number along the $f$ series of the elements in the periodic table. The element Np, situated between U and Pu, seems to be at the localized-itinerant border.
Recent NMR results on NpCoGa$_5$ suggest localized behavior in contrast to the more itinerant-like UCoGa$_5$\cite{kambe}.
Among the Np compounds, NpO$_2$ is an example where the localized description works well to describe the triple-{\bf q} octupolar order below $25$K. Even in the case of U-based compounds, the localized picture may apply to the case of the famous hidden ordered phase of URu$_2$Si$_2$.
Although our simple CEF model can explain
the diverse behavior realized in NpTGa$_5$,
the purely localized model cannot explain the enhanced $T$-linear term in the specific heat, for
example.
Hence a more sophisticated model is desirable to account for the dual character of $5f$
electrons.  We note that in the case of even number of $5f$ electrons as in Np$^{3+}$,
electronic states in the localized limit and the band limit may be connected continuously \cite
{watanabe}.

A further question is the value of the ordered magnetic moments at low temperatures. It is found experimentally that in the case of NpCoGa$_5$ the saturated magnetic moment is about $0.84\mu_{\rm B}$/Np, and it is $0.96\mu_{\rm B}$/Np in NpRhGa$_5$. In the mean field theory, the zero temperature value of the moments is mainly determined by the crystal field parameters. For example, with crystal field parameter $a=0.87$ we get $1.22\mu_{\rm B}$ for the saturated magnetic moment $J_z$ in the case of NpCoGa$_5$, which is larger than the observed value. We saw previously in the case of NpRhGa$_5$ that the presence of quadrupolar interaction $\lambda_{xy}$ can reduce the value of the magnetic moment $\langle J_x+J_y\rangle$.
On the other hand, quantum fluctuations can lead to the reduction of the ordered moment even for localized electrons. 
We expect stronger fluctuations in the Np $115$ systems because of the strong two-dimensional feature compared to the case of the cubic NpGa$_3$, for example, in which the ordered moments are about $1.51\mu_{\rm B}$\cite{npga3}.

The main properties of NpTGa$_5$ systems with T=Co, Ni and Rh can be understood within the two-dimensional model.
However, in order to discuss behavior in the presence of magnetic field like susceptibility anisotropy, for example, the interlayer coupling should be also included. Only in the case of NpCoGa$_5$ 
we have included an interlayer interaction, and calculated properties in external magnetic field. It would be interesting to incorporate interlayer coupling also in the cases of T=Ni and Rh and compare the obtained results to the measured ones. 
The difference in the number of conduction electrons may be a reason why
the interlayer RKKY coupling becomes ferromagnetic in NpNiGa5, as compared with observed antiferromagnetic interlayer ordering in 
NpCoGa$_5$ and NpRhGa$_5$ 
We plan to discuss these issues in a subsequent work.

\section*{Acknowledgment}

The authors are grateful to
F. Honda,  S. Jonen,  D. Aoki, K. Kaneko, S. Kambe and N. Metoki
for showing their experimental results on NpTGa$_5$ prior to publication,
and for useful correspondence.


\begin{thebibliography}{99}


\bibitem{pu}
J. L. Sarrao, L. A. Morales, J. D. Thompson, B. L. Scott, G. R. Stewart, F. Wastin, J. Rebizant, P. Boulet, E. Colineau and G. H. Lander: Nature {\bf 420} (2002) 297.

\bibitem{petrovich}
C. Petrovich, P. G. Pagliuso, M. F. Hundley, R. Movshovich, J. L. Sarrao, J. D. Thompson, Z. Fisk and P. Monthoux: J. Phys.: Condensed Matter {\bf 13} (2001)
L337.


\bibitem{aoki}
D. Aoki, Y. Homma, Y. Shiokawa, E. Yamamoto, A. Nakamura, Y. Haga, R. Settai, T. Takeuchi and Y. $\bar{{\rm O}}$nuki: J. Phys. Soc. Jpn. {\bf 73} (2004) 1665.

\bibitem{metoki}
N. Metoki, K. Kaneko, E. Colineau, P. Javorsk\'y, D. Aoki, Y. Homma, P. Boulet, F. Wastin, Y. Shiokawa, N. Bernhoeft, E. Yamamoto, Y. $\bar{{\rm O}}$nuki, J. Rebizant and G. H. Lander: Phys. Rev. B {\bf 72} (2005) 014460.

\bibitem{colineau}
E. Colineau, P. Javorsk\'y, P. Boulet, F. Wastin, J. C. Griveau, J. Rebizant, J. P. Sanchez and G. R. Stewart: Phys. Rev. B {\bf 69} (2004) 184411.

\bibitem{aoki2}
D. Aoki et al, Y. Homma, Y. Shiokawa, H. Sakai, E. Yamamoto, A. Nakamura, Y. Haga, R. Settai and Y. $\bar{{\rm O}}$nuki: J. Phys. Soc. Jpn. {\bf 74} (2005) 2323.


\bibitem{jonen}
S. Jonen et al., submitted to PRB.

\bibitem{colineau2}
E. Colineau, F. Wastin, P. Boulet, P. Javorsk\'y, J. Rebizant and J. P. Sanchez: J. Alloys and Compounds {\bf 386} (2005) 57.

\bibitem{honda}
F. Honda, N. Metoki, K. Kaneko, D. Aoki, Y. Homma, E. Yamamoto, Y. Shiokawa, Y. $\bar{{\rm O}}$nuki, E. Colineau, N. Bernhoeft and G. H. Lander: Physica B {\bf 359-361} (2005) 1147.

\bibitem{honda2}
F. Honda et al., to be published.


\bibitem{kuramoto1}
H. Kusunose and Y. Kuramoto, J. Phys. Soc. Jpn. {\bf 70} (2001) 1751.

\bibitem{kuramoto3}
Y. Kuramoto and N. Fukushima, J. Phys. Soc. Jpn. {\bf 67} (2000) 583.

\bibitem{kuramoto2}
Y. Kuramoto and H. Kusunose, J. Phys. Soc. Jpn. {\bf 69} (2000) 671.


\bibitem{libero}
V. L. Libero and D. L. Cox, Phys. Rev. B {\bf 48} (1993) 3783.


\bibitem{nppt}
F. Honda et al., to be published.



\bibitem{kambe}
S. Kambe et al., to be published.



\bibitem{watanabe}
S.  Watanabe, Y. Kuramoto, T. Nishino and N. Shibata,
J. Phys. Soc. Jpn. {\bf 68} (1999) 159.

\bibitem{npga3}
M. N. Bouillet, T. Charvolin, A. Blaise, P. Burlet, J. M. Fournier, J. Larroque and J. P. Sanchez: J. Magn. Magn. Mater. {\bf 125} (1993) 113.

\end{thebibliography}
\end{document}